\documentstyle[12pt]{article}
\newcommand{\be}{\begin{equation}}
\newcommand{\ee}{\end{equation}}
\newcommand{\ben}{\begin{eqnarray}\displaystyle}
\newcommand{\een}{\end{eqnarray}}
\newcommand{\refb}[1]{(\ref{#1})}

\begin{document}

{}~ \hfill\vbox{\hbox{hep-th/9608005}\hbox{MRI-PHY/96-23}}\break

\vskip 3.5cm

\centerline{\large \bf BPS States on a Three Brane Probe}

\vspace*{6.0ex}

\centerline{\large \rm Ashoke Sen\footnote{On leave of absence from 
Tata Institute of Fundamental Research, Homi Bhabha Road, 
Bombay 400005, INDIA}
\footnote{E-mail: sen@mri.ernet.in, sen@theory.tifr.res.in}}

\vspace*{1.5ex}

\centerline{\large \it Mehta Research Institute of Mathematics}
 \centerline{\large \it and Mathematical Physics}

\centerline{\large \it 10 Kasturba Gandhi Marg, 
Allahabad 211002, INDIA}

\vspace*{4.5ex}

\centerline {\bf Abstract}

Recently Banks, Douglas and Seiberg
have shown that the world volume theory of
a three brane of the type IIB theory in the presence of a configuration 
of four Dirichlet seven branes and an
orientifold plane is described by an N=2 supersymmetric SU(2) gauge theory
with four quark flavours 
in 3+1 dimensions. In this note we show how the BPS mass formula for
N=2 supersymmetric gauge theory arises from masses of open strings
stretched between the three brane and the seven brane along 
appropriate geodesics.

\vfill \eject

\baselineskip=18pt

Recent work by Banks, Douglas and Seiberg has shown that the world-volume
theory of a three brane, moving in the background of an orientifold
seven plane and a set of four Dirichlet seven branes is described by
an N=2 supersymmetric SU(2) gauge theory in 3+1 dimensions with four
hypermultiplets in the fundamental representation of SU(2)\cite{BDS}. 
There is a dual description of this background\cite{AS} known 
as F-theory\cite{FTHEORY}, where the three brane moves in the
background of six seven branes. In the weak coupling limit,
four of these seven branes can be regarded as 
conventional D-branes, and the other two are related to it by 
SL(2,Z) S-duality transformations of the type IIB theory. In both 
descriptions the moduli space is spanned by the complex coordinate
of the three brane in the plane transverse to the seven-branes.
The first
description represents the semiclassical limit of the moduli space of
N=2 supersymmetric SU(2) gauge theory with four quark flavours, with the
orientifold plane representing the point where SU(2) gauge symmetry is
restored semiclassically, and the locations of the four Dirichlet
seven branes representing the points where the hypermultiplets become
massless. On the other hand the F-theory description gives the full
quantum corrected moduli space of the N=2 supersymmetric SU(2) gauge
theory\cite{SW}, with the six seven branes representing the 
locations in the
moduli space where one of the four elementary hypermultiplets, the
monopole or the dyon becomes massless. 

Given the fact that the three brane world volume theory is an N=2 
supersymmetric SU(2) gauge theory with four hypermultiplets, one can
ask: what are the analogs of the BPS states on the three brane world
volume? Qualitatively, the answer to this question is simple, $-$ these
are just the open string states stretched between the three brane and
a seven brane.\footnote{There are also BPS states corresponding to
open strings starting and ending on the
three brane. These correspond to massive gauge bosons and states that are 
related to them via SL(2,Z) transformation, 
and can be treated in a similar manner, but we shall not discuss them
further in this paper.}\footnote{The possibility of representing these
BPS states as stretched (self-dual) strings has also  been discussed in
ref.\cite{LEWA}; however there the string stretches along the Riemann
surface associated with the Seiberg-Witten curve instead of along the
moduli space.}
To be more specific, let us define a $(p,q)$ string as
the bound state of $p$ elementary type IIB strings and $q$ 
D-strings\cite{WITTBOUND}. We shall also define a $(p,q)$ seven-brane
to be the seven-brane on which a $(p,q)$ string can end\cite{DOLI}. 
Thus, for
example an (1,0) string is just an elementary type IIB string, and an
(1,0) seven brane is just the Dirichlet seven brane of type IIB theory.
More general $(p,q)$ strings or seven-branes are obtained from (1,0)
strings / seven-branes via S-duality transformation. With this 
convention the F-theory background under consideration in the weak
coupling limit can be taken to consist of
four (1,0) seven branes, one (0,1) seven brane and one (2,1) seven brane.
Note however that there is non-trivial SL(2,Z) monodromy as we go around
the various seven branes, and so when `viewing' a seven brane from a 
fixed point (say at
infinity) we can find complicated paths, along which a (1,0) (or (0,1)
or (2,1)) seven-brane may appear as a $(p,q)$ seven brane.

While a particular type of seven-brane can only have a particular kind
of string ending on it, the three brane in the type IIB theory can have
any $(p,q)$ string ending on it. This is related to the fact that the
three brane is invariant under an SL(2,Z) transformation. Thus we can
have a state represented by a $(p,q)$ string starting on the three brane
and ending on a $(p,q)$ seven brane.\footnote{In principle for any pair
$(p,q)$ relatively prime, we can find paths along which a given seven brane
will appear as a $(p,q)$ seven brane to the three brane, and the $(p,q)$
open string can stretch along such a path. However,
in general not all of them will
represent stable BPS states.} From the three brane world-volume
point of view this will correspond to a state carrying $p$ units of
electric charge and $q$ units of magnetic charge\cite{GREE}. 
We shall refer to this
state as a $(p,q)$ state. As the three brane approaches the location of
a $(p,q)$ seven-brane, the open string stretched between the three brane
and the seven brane becomes massless. Thus the location of a $(p,q)$
7-brane can be interpreted as the point in the moduli space where a
$(p,q)$ state becomes massless. Note however that due to non-trivial
SL(2,Z) monodromy in the moduli space, the values
of $(p,q)$ associated with such a point depend on the path in the
moduli space that we take to approach such a point.

We shall be interested in deriving the mass of a $(p,q)$ state at a
generic point in the moduli space when the
three brane is away from the location of the $(p,q)$ seven-brane. 
According to ref.\cite{SW} it is given by
\be \label{e1}
m_{p,q}^{SW}=|p a(z) + q a_D(z) +\sum_i m_i S_i|\, ,
\ee
where $a$ and $a_D$ are known functions of the gauge invariant coordinate
$z$ (called $u$ in \cite{SW}) labelling the moduli space of the $3+1$
dimensional gauge theory;
$m_i$ are the bare masses of the hypermultiplets, and
$S_i$ are the global U(1) charges carried by the state. 
In the present context $z$ can be identified
to the complex coordinate of the three brane in the plane transverse to
the seven-branes and $m_i^2$ can be identified as the positions of the
four Dirichlet seven branes in the orientifold description\cite{AS,BDS}. 
We shall use
a slightly different form of eq.\refb{e1}. Let us focus on a $(p,q)$
state carrying a given set of global U(1) charges.
If $z_0$ denotes the point
on the moduli space where the particular $(p,q)$ state under 
consideration becomes massless, then $m_{p,q}^{SW}$ must vanish at 
$z_0$.
This gives,
\be \label{enew}
pa(z_0)+q a_D(z_0) + \sum_i m_i S_i=0\, .
\ee
Hence eq.\refb{e1} can be rewritten as
\be \label{e2}
m_{p,q}^{SW}=|p a(z) + q a_D(z) - pa(z_0)-qa_D(z_0)|\, .
\ee

We shall show that this expression can be reinterpreted as the mass
of a $(p,q)$ string stretched between the three brane and the
$(p,q)$ seven brane. To this end note that the string tension of a
$(p,q)$ string (measured in the canonical metric) is given 
by\cite{SCHWARZ}
\be \label{e3}
T_{p,q}={1 \over \sqrt{\lambda_2}}|p+q\lambda|\, ,
\ee
where
\be \label{e4}
\lambda= a + i e^{-\Phi/2}\equiv \lambda_1 + i\lambda_2\, ,
\ee
$a$ being the massless scalar arising from the Ramond-Ramond sector of the
type IIB theory and $\Phi$ being the dilaton field. Thus the mass of a
$(p,q)$ string state stretched along a curve $C$ will be given by
\be \label{e5}
\int_C T_{p,q} ds\, ,
\ee
where $ds$ is the line element along the curve $C$ measured in the
canonical metric. 

Let us now apply this formula to the specific $F$-theory background that we
have been considering. 
We shall follow the notation of ref.\cite{AS}.
The $F$-theory background is given by\cite{AS}
\ben \label{e6}
\lambda(z) & = & \tau(z) \equiv \tau_1+i\tau_2\, , \nonumber \\
ds^2 & = & \tau_2 \eta(\tau)^2\bar\eta(\bar\tau)^2 
\prod_{i=1}^6(z-z_i)^{-1/12}
(\bar z-\bar z_i)^{-1/12} dz d\bar z\, ,
\een
where $\tau$ is determined from the equation
\be \label{e7}
j(\tau)={4. (24f)^3\over 4f^3+27 g^2}\, ,
\ee
$f$ and $g$ being arbitrary
polynomials in $z$ of degree 2 and 3 respectively. $z_i$
represent the six zeroes of
\be \label{e8}
\Delta=4f^3+27 g^2\,  
\ee
where $\tau\to i\infty$ up to an SL(2,Z) transformation. Here $\eta(\tau)$ 
represents the Dedekind eta function, and $j(\tau)$ is the
modular function of $\tau$ with a single pole at $\tau=i\infty$ and zero
at $\tau=e^{i\pi/3}$.

Using eqs.\refb{e3}, \refb{e5} and \refb{e6} we see that the mass 
of a $(p,q)$ string
stretched between a $(p,q)$ seven-brane located at $z_0$ and a three brane
located at $z$ along a curve $C$ is given by
\be \label{e9}
m_{p,q}=\int_C
|\eta(\tau)^2\prod_{i=1}^6(z-z_i)^{-1/12}(p+q\tau)\, dz|\, .
\ee
Introducing a new coordinate $w_{p,q}$ through the relation
\be \label{e10}
dw_{p,q}=
\eta(\tau)^2\prod_{i=1}^6(z-z_i)^{-1/12}(p+q\tau)\, dz\, ,
\ee
we can rewrite eq.\refb{e9} as
\be \label{e10a}
m_{p,q} = \int_C|dw_{p,q}|\, .
\ee
In order to get a BPS state, we need to choose $C$ so as to
minimize the mass of the open string stretched between $z_0$ and $z$.
By looking at eq.\refb{e10a} it is clear that in the $w_{p,q}$ coordinate
system this corresponds to the
straight line from the point $w_{p,q}(z_0)$ to $w_{p,q}(z)$. Thus the 
mass of the BPS
state represented by the $(p,q)$ open string stretched between the point
$z_0$ and $z$ is given by
\be \label{e11}
m_{p,q}^{BPS} = |w_{p,q}(z) -w_{p,q}(z_0)|\, .
\ee

We now want to compare this with eq.\refb{e2}. To do this we 
use the equation (which we shall prove later)
\be \label{e12}
da=
\eta(\tau)^2\prod_{i=1}^6(z-z_i)^{-1/12}dz\, .
\ee
In that case, using eq.\refb{e10} and the fact that\cite{SW}
\be \label{e13}
da_D=\tau da\, ,
\ee
we get
\be \label{e14}
dw_{p,q}= p da+q da_D\, .
\ee
Eq.\refb{e11} can now be rewritten as
\be \label{e15}
m_{p,q}^{BPS}=|pa(z)+qa_D(z)-pa(z_0)-qa_D(z_0)|\, .
\ee
This agrees with the BPS formula \refb{e2} for N=2 supersymmetric SU(2)
gauge theory derived by Seiberg and Witten.

Thus it remains to prove eq.\refb{e12}. Let us define
\be \label{e16}
F(z)=
{da\over dz}
\eta(\tau)^{-2}\prod_{i=1}^6(z-z_i)^{1/12}\, .
\ee
We shall show that $F(z)$ has no singularity in the $z$ plane and 
asymptotically goes to a constant (which can be set to unity by a rescaling
of $z$). This would prove that $F(z)=1$ and hence establish eq.\refb{e12}.
The possible singularities of $F$ can arise at one of the $z_i$'s. Let us
focus on one of them, say $z_1$, and let us suppose it denotes the
location of an $(m,n)$ seven brane in the $z$ plane. 
Then near $z_1$ we get\cite{SW},
\ben \label{e17}
m a + n a_D & \simeq & c (z-z_1)\, , \nonumber \\
r a + s a_D & \simeq & {c\over 2\pi i} (z-z_1)\ln (z-z_1) + 
\hbox{constant} \, ,\nonumber \\
{s\tau + r\over n\tau+m} & \simeq & {1\over 2\pi i} \ln (z-z_1)\, ,
\een
where $\pmatrix{m & n\cr r & s}$ is an SL(2,Z) matrix and $c$ is a constant. 
This gives
\be \label{e18}
a\simeq -{c \over 2\pi i} n (z-z_1) \ln(z-z_1) + \hbox{constant} \, ,
\ee
so that
\be \label{e19}
{da\over dz} \simeq -{c \over 2\pi i} n \ln(z-z_1)\, .
\ee
We also get,
\be \label{e20}
\eta^2(\tau) \simeq (z-z_1)^{1/12} {n\over 2\pi i} \ln (z-z_1)\, 
\ee
up to a phase.
Substituting this asymptotic behaviour of $(da/dz)$ and $\eta$ in 
the right hand side of eq.\refb{e16} we see that
$F(z)$ is non-singular at $z=z_1$. Similarly $F(z)$ can be shown to 
be non-singular near the other $z_i$'s as well. Finally, as $z\to \infty$,
$a\propto \sqrt{z}$\cite{SW}, and $\tau$ approaches a constant; hence $F(z)$
approaches a constant.
This shows that $F(z)$ must be a
constant independent of $z$, thereby proving the desired result.

To summarize, we have derived the mass  formula for the BPS states in N=2
supersymmetric gauge theory by examining the masses of open strings 
stretched between the three brane and the seven brane for the configuration
studied in ref.\cite{BDS}. This result is likely to 
be relevant in the study of
a three brane dynamics in the presence of more complicated seven brane
configurations, notably those which correspond to the presence of more
than one orientifold plane. Representing BPS states as stretched open
strings should also be useful in determining the
full spectrum of BPS states at a given point in the moduli space, a question
that has already been addressed by other methods before\cite{SW,BIFE,LEWA}.

\noindent{\bf Acknowledgement}:
I wish to thank S.Mukhi and C. Vafa for useful discussions.
I also wish to acknowledge the hospitality of LPTHE at University of
Pierre and Marie Curie during the course of this work.

\end{document}